\documentclass[12pt]{article}
\usepackage{graphicx}

\begin{document}

{\Large \bf Analytic derivation of an approximate SU(3) symmetry inside the symmetry triangle of the Interacting Boson Approximation model} 

\bigskip 

\centerline{Dennis Bonatsos$^1$, S. Karampagia$^1$, and R.F. Casten$^2$}

\medskip 

\centerline{$^1$Institute of Nuclear Physics, National Centre for Scientific Research 
``Demokritos'',}

\centerline{ GR-15310 Aghia Paraskevi, Attiki, Greece}

\centerline{$^2$ Wright Nuclear Structure Laboratory, Yale
University, New Haven, CT 06520, USA}

\vskip 0.5truein

\centerline{\bf Abstract}

Using the contraction of the SU(3) algebra to the algebra of the rigid rotator in the large boson number limit 
of the Interacting Boson Approximation (IBA) model, a line is found inside the symmetry triangle of the IBA, 
along which the SU(3) symmetry is preserved. The line extends from the SU(3) vertex to near the critical line
of the first order shape/phase transition separating the spherical and prolate deformed phases, and
lies within the Alhassid--Whelan arc of regularity, the unique valley of regularity connecting the SU(3) and U(5) 
vertices amidst chaotic regions. In addition to providing an explanation for the existence of the arc of regularity,
the present line represents the first example of an analytically determined approximate symmetry in the interior 
of the symmetry triangle of the IBA. The method is applicable to algebraic models possessing subalgebras amenable 
to contraction. This condition is equivalent to algebras in which the equilibrium ground state (and its rotational band) 
become energetically isolated from intrinsic excitations, as typified by deformed solutions to the IBA 
for large numbers of valence nucleons. 

\bigskip



\section{Introduction} 

The study of chaotic properties of the Interacting Boson Approximation (IBA) model \cite{IA}, both classically 
and quantum mechanically, led to the discovery \cite{AWPRL,AWNPA} 
of a narrow strip of nearly regular behavior
inside the symmetry triangle \cite{triangle} of the IBA, connecting the U(5) and SU(3) limiting symmetries,  
in addition to the regular region along the U(5)-O(6) leg of the triangle. 
While the existence of the latter is known to be due to the underlying SO(5) symmetry, 
a common subalgebra of both U(5) and O(6) present throughout the U(5)-O(6) leg \cite{Talmi},
the origin of the former, called the Alhassid--Whelan (AW) arc of regularity,  has remained 
an open question. 

The existence of a nearly regular region 
connecting U(5) and SU(3), has been corroborated by studies of the wave-function entropy in the IBA 
\cite{CJPRE}. From the empirical point of view, it has been found \cite{AWexp} that the line 
corresponding to the degeneracy of the $\beta_1$ and $\gamma_1$ bandheads ($2_{\gamma_1}^+=0_{\beta_1}^+$)
closely follows the arc of regularity, and 12 nuclei
closely exhibiting this behavior have been located. Recently, it has been realized \cite{Macek75} that the locus 
of the $2_{\gamma_1}^+=0_{\beta_1}^+$ degeneracy closely follows the line of change of stability  
of $\gamma$-vibrations at low energies. 

The presence of (near) regularity presupposes the existence of some underlying (approximate) symmetry.
The degeneracy of the $\beta_1$ and $\gamma_1$ bands is a well known hallmark of the SU(3) symmetry 
of the IBA \cite{IA}. It has recently been found \cite{arcPRL} that 
imposing the $2_{\gamma_1}^+=2_{\beta_1}^+$ degeneracy in the IBA framework
leads to a line inside the symmetry triangle of the IBA which, 
in the region between the SU(3) vertex and the shape/phase coexistence region \cite{IZC}
(separating spherical from prolate deformed shapes), 
is located very close to the arc of regularity, while 
at the same time the degeneracies predicted by SU(3), not only for the $\beta_1$ and $\gamma_1$ bands, 
but also for bands belonging to higher irreducible representations (irreps) of SU(3), are preserved 
to a very good extent \cite{arcPRL}. This result extends the notion of quasidynamical symmetry (QDS), originally 
introduced \cite{Rowe1225,Rowe2325,Rowe745,Rowe756,Rowe759}
for describing the persistence of limiting symmetries along the U(5)-O(6) and 
U(5)-SU(3) legs of the IBA, to the interior of the triangle. 
The domain of validity of the SU(3) QDS inside the IBA symmetry triangle 
has also been considered by mean field techniques \cite{Macek80}.
The analysis of Ref. \cite{arcPRL} was limited 
to the low-lying part of the spectrum, while regularity amidst chaoticity has been discovered 
by Alhassid and Whelan through the study of the whole spectrum \cite{AWPRL,AWNPA}. 
The properties of high-lying rotational bands built on axially deformed ground states 
have been studied recently \cite{Macek82} in the IBA, 
showing signatures of a SU(3) QDS extending to the highest part of the IBA spectrum. 

In the present work, we study Hamiltonians that, in the large boson number limit of the IBA, 
approximately commute with the SU(3) generators. In addition, we derive an analytic expression for the locus
of these Hamiltonians inside the IBA symmetry triangle.  
Moreover, this locus in fact corresponds closely to the arc of regularity.
  
This represents the first {\sl analytical} identification of an approximate symmetry {\sl within} the 
symmetry triangle of the IBA. The proof takes advantage of the well known contraction \cite{Inonu} of the SU(3) algebra to the 
[R$^5$]SO(3) algebra \cite{LeBlanc,RowePPNP} of the rigid rotator \cite{Ui}. 
Furthermore, using the contraction of O(6) to the [R$^5$]SO(5) algebra \cite{MtV84,Elliott171} of the $\gamma$-unstable rotator,
we prove that no line related to the O(6) symmetry exists within the triangle.  

The IBA Hamiltonian used is described in Section 2, while in Section 3 the equation of the line 
corresponding to the SU(3) symmetry is derived. The $\overline{\rm SU(3)}$, O(6), and O(5) symmetries 
are considered in Sections 4--6, while an alternative parametrization is presented in Section 7. The conclusions and outlook 
are presented in Section 8. The commutation relations and matrix elements needed for the derivations are given explicitly in Appendices 1 and 3, 
while the details of the SU(3)$\to$[R$^5$]SO(3) and O(6)$\to$[R$^5$]SO(5) contractions are given in Appendices 2 and 4 
respectively.  

\section{The IBA Hamiltonian and symmetry triangle} 

The IBA Hamiltonian used by Alhassid and Whelan \cite{AWPRL,AWNPA} reads  
\begin{equation}\label{HAW}
\hat H(\eta,\chi) = \hat H_1 + \hat H_2 =
c \left[ \eta \hat n_d +{\eta- 1 \over N}
\hat Q^{(2)}_\chi \cdot \hat Q^{(2)}_\chi\right],
\end{equation}
where $N$ is the number of valence bosons, $c$ is a
scaling factor, $\hat H_1$ and $\hat H_2$ denote the first and the second term respectively, 
\begin{equation} 
\hat n_d = d^\dagger \cdot \tilde d = \sqrt{5} (d^\dagger \tilde d)^{(0)},
\end{equation}   
\begin{equation} 
\hat Q^{(2)}_{\chi,\xi} = (s^\dagger \tilde d + d^\dagger s)^{(2)}_\xi +\chi (d^\dagger \tilde d)^{(2)}_\xi.
\end{equation}
The above Hamiltonian contains two parameters,
$\eta$ and $\chi$, with the parameter $\eta$ ranging from 0 to
1, and the parameter $\chi$ ranging from 0 to $-\sqrt{7}/2=-1.323$.
The U(5) symmetry limit corresponds to $\eta=1$, the SU(3) limit 
to $\eta=0$, $\chi=-\sqrt{7}/2$, and the O(6) limit to $\eta=0$, $\chi=0$. 
These symmetries are placed at the vertices of the symmetry triangle \cite{triangle} of the IBA, 
shown in Fig.~1(a). In the symmetry triangle, the narrow coexistence region \cite{IZC} surrounding 
the critical line \cite{Werner} separating the spherical phase from the prolate deformed phase, corresponding 
to a first-order shape/phase transition \cite{Deans}, is also shown. It corresponds to $\eta \sim 0.8$. 
The point at which the critical line reaches 
the U(5)-O(6) side of the triangle is known to correspond to a second-order shape/phase transition \cite{Deans}. 

Note that an alternate parametrization of the Hamiltonian of Eq. (\ref{HAW}) is often used. 
We will discuss the present results in the context of that parametrization in Section 7.

\section{The SU(3) symmetry} 

\subsection{Commutation relations} 
The SU(3) \cite{IA} algebra is generated by the angular momentum operators 
\begin{equation}\label{L} 
\hat L_\xi = \sqrt{10} (d^\dagger \tilde d)^{1}_\xi,
\end{equation}
and the quadrupole operators 
\begin{equation} 
\hat Q^{(2)}_{SU(3),\xi} = (s^\dagger \tilde d + d^\dagger s)^{(2)}_\xi -{\sqrt{7}\over 2} (d^\dagger \tilde d)^{(2)}_\xi.
\end{equation}

In order to have an underlying SU(3) symmetry, the Hamiltonian of Eq. (\ref{HAW}) has to commute with the generators of SU(3).
It does commute with the angular momentum operators $\hat L_\xi$ by construction, since it is a scalar quantity.
We will examine the special conditions under which the Hamiltonian also commutes (approximately) with the quadrupole operators.  
The commutation relations needed for this task are listed in Appendix~1. 

The first term of the Hamiltonian gives 
\begin{equation}\label{term1}
[\hat H_1, \hat Q^{(2)}_{SU(3),\nu}] = c \eta [\hat n_d, \hat Q^{(2)}_{SU(3),\nu}] = c \eta (d^\dagger s -s^\dagger \tilde d)^{(2)}_\nu.
\end{equation}

Using 
\begin{equation}\label{QSU3}
\hat Q ^{(2)}_{\chi,\xi} = \hat Q ^{(2)}_{SU(3),\xi} + \left(\chi + {\sqrt{7} \over 2} \right) (d^\dagger \tilde d)^{(2)}_\xi ,  
\end{equation}
in the second term of the Hamiltonian one gets the intermediate result
$$
[\hat Q^{(2)}_\chi \cdot \hat Q^{(2)}_\chi, \hat Q^{(2)}_{SU(3),\nu}] = 
\sum_\xi (-1)^\xi \left\{ [\hat Q^{(2)}_{SU(3),\xi}, \hat Q^{(2)}_{SU(3),\nu}] \hat Q^{(2)}_{\chi, -\xi} + 
\hat Q^{(2)}_{\chi, \xi} [\hat Q^{(2)}_{SU(3),-\xi}, \hat Q^{(2)}_{SU(3),\nu}] \right.$$
\begin{equation}\label{interm} \left.
+ \left( \chi + {\sqrt{7}\over 2} \right) \left\{ [ (d^\dagger \tilde d)^{(2)}_\xi , \hat Q^{(2)}_{SU(3), \nu} ] \hat Q^{(2)}_{\chi, -\xi}
+ \hat Q^{(2)}_{\chi, \xi} [ (d^\dagger \tilde d)^{(2)}_{-\xi} , \hat Q^{(2)}_{SU(3), \nu} ]
\right\} \right\} . 
\end{equation}

In order to obtain the conditions for which the Hamiltonian of Eq. (\ref{HAW}) commutes with the generators of SU(3), 
we exploit a simplification of Eq. (\ref{interm}) that occurs in the large $N$ limit. 
In this limit, the eigenvalue expression for the second order Casimir of SU(3) [see Eq. (\ref{SU3c})] 
reduces to just the $\lambda ^2$ term for SU(3) irreducible representations (irreps) $(\lambda,\mu)$ with $\lambda >> \mu$ 
and hence the ground state band [which belongs to the (2N,0) irrep] becomes energetically isolated from all other excitations.  
That is, SU(3) effectively reduces to a simple rigid rotator. This situation is formally known as the contraction of SU(3) 
to R$^5$[SO(3)] \cite{LeBlanc,RowePPNP} and occurs when the $Q^{(2)}_{SU(3)}$ operators can be replaced 
by mutually commuting quantities [For a detailed explanation, see in Appendix 2 the discussion leading to Eq. (\ref{qqSU3b})]. 
If the $Q^{(2)}_{SU(3)}$ operators can be approximated by mutually commuting quantities,
Eq. (\ref{interm}) greatly simplifies. These results are discussed in detail in Appendix 2.

In the large $N$ limit, where contraction occurs,  
the commutators in the first two terms in Eq. (\ref{interm}) will vanish. Furthermore,  the vanishing of the commutator 
\begin{equation}
[ \hat Q^{(2)}_{SU(3),\xi},  \hat Q^{(2)}_{SU(3),\nu}] = {15\over 4} (2 \xi 2 \nu | 1 \xi+\nu) (d^\dagger \tilde d)^{(1)}_{\xi+\nu}  
\end{equation}
in this limit, implies that terms containing $(d^\dagger \tilde d)^{(1)}$ can be ignored.
This fact can be understood qualitatively as a consequence of the relevant dominance of $s$ bosons over $d$ bosons
within the ground state band, especially for relatively low-lying states in the large boson number limit. 

Eq. (\ref{interm}) can be rewritten, without using any approximations yet, as 
$$
[\hat Q^{(2)}_\chi \cdot \hat Q^{(2)}_\chi, \hat Q^{(2)}_{SU(3),\nu}] = {3\sqrt{15}\over 4}  
[ ((d^\dagger \tilde d)^{(1)} \hat Q^{(2)}_{\chi})^{(2)} _\nu -
(\hat Q^{(2)}_{\chi}  (d^\dagger \tilde d)^{(1)} )^{(2)}_\nu ] $$
$$
+ \left( \chi + {\sqrt{7} \over 2} \right)  [((d^\dagger s- s^\dagger \tilde d)^{(2)} \hat Q^{(2)}_\chi )^{(2)}_\nu 
+ ( \hat Q^{(2)}_\chi   (d^\dagger s -s^\dagger \tilde d)^{(2)} )^{(2)}_\nu ] $$
\begin{equation}\label{interm4} 
+ \left( \chi + {\sqrt{7} \over 2} \right) \sum_{k=1,3} \sqrt{35(2k+1)} 
\left\{ \matrix{2 & 2 & k \cr 2 & 2 & 2 \cr} \right\} 
[((d^\dagger \tilde d )^{(k)} \hat Q^{(2)}_\chi )^{(2)}_\nu 
- ( \hat Q^{(2)}_\chi   (d^\dagger \tilde d )^{(k)})^{(2)}_\nu ].  
\end{equation} 
In the large $N$ limit the terms containing $(d^\dagger \tilde d)^{(k)}$ (in the first line and in the third line) can be omitted.
Furthermore, in the second line, $\hat Q^{(2)}_\chi$ can be replaced by $\hat Q^{(2)}_{SU(3)}$, since, as seen from Eq. (\ref{QSU3}), 
they differ by terms $(d^\dagger \tilde d)^{(2)}$, which are small. 
In addition, in this limit $\hat Q^{(2)}_{SU(3)}$ can be replaced by the intrinsic quadrupole moment (a scalar),
which is $N \sqrt{2}$ in the present case (see Appendix A2). This replacement will be justified in detail in subsection 3.2~.  
This result is perhaps familiar in the context of the well-known property of SU(3) 
that $B(E2: 2_1^+ \to 0_1^+)$ goes as $N^2$ \cite{IA}, that is, the collectivity of yrast transition strengths 
increases quadratically with boson number.  
Then in the large $N$ limit one is left with 
\begin{equation}\label{inter5}
[\hat Q^{(2)}_\chi \cdot \hat Q^{(2)}_\chi, \hat Q^{(2)}_{SU(3),\nu}] = 2 \sqrt{2} N \left( \chi + {\sqrt{7} \over 2} \right) 
 (d^\dagger s -s^\dagger \tilde d)^{(2)}_\nu.
\end{equation}
Then in the large $N$ limit the commutator for the second part of the Hamiltonian reads
\begin{equation}\label{term2}
[\hat H_2, \hat Q^{(2)}_{SU(3),\nu}]= c (\eta-1) 2\sqrt{2} \left( \chi + {\sqrt{7} \over 2} \right) (d^\dagger s -s^\dagger \tilde d)^{(2)}_\nu.
\end{equation}
In order to get a vanishing commutator, the coefficients of $ (d^\dagger s -s^\dagger \tilde d)^{(2)}_\nu $ in Eqs. (\ref{term1}) 
and (\ref{term2}) should cancel, leading in the large $N$ limit to the condition 
\begin{equation}\label{line_eta}
\chi(\eta) = {1\over 2\sqrt{2}} {\eta \over (1-\eta)} - {\sqrt{7}\over 2}.
\end{equation}
When $\chi$ is taking values between $-\sqrt{7}/2$ and 0, the parameter $\eta$ takes values between 
1 and 0.789~. From the formulae reported in Refs. \cite{AWexp} and \cite{Libby} it is clear that the critical line 
in the large $N$ limit corresponds to $\eta_{crit}=0.8$ for $\chi=0$ and to $\eta_{crit}=9/11=0.818$ for $\chi=-\sqrt{7}/2$.
Thus the line described by Eq. (\ref{line_eta}) cannot reach the critical line, confined in the region between the critical line 
and the SU(3) vertex.  

It should be noticed that the arc of regularity found in Refs. \cite{AWPRL,AWNPA} 
has been approximately described by \cite{CJPRE}
\begin{equation}\label{arc} 
\chi(\eta)= {\sqrt{7}-1 \over 2} \eta - {\sqrt{7} \over 2}. 
\end{equation}

The similarity between the lines described by Eqs. (\ref{line_eta}) and (\ref{arc}) can be seen in Fig.~2.
Indeed, the two equations give very similar predictions for values of $\eta$ between 0 and 0.6, i.e., 
from the SU(3) vertex until quite close to the critical line. 
 
The symmetry triangle of IBA in the Alhassid--Whelan parametrization is shown in Fig.~1(a),
together with the arc corresponding to Eq. (\ref{arc}) and the line of Eq. (\ref{line_eta}). 
The degeneracy line corresponding to $E(2^+_\beta)=E(2^+_\gamma)$, found in Ref. \cite{arcPRL},
is also shown (on the right of the critical line) for comparison.

We see that the present line remains very close to both the $E(2^+_\beta)=E(2^+_\gamma)$ degeneracy line
and the original arc line from the SU(3) vertex until quite close to the critical region, where 
both the $E(2^+_\beta)=E(2^+_\gamma)$ degeneracy line and the present line turn upwards, avoiding 
to meet the critical line. 

In Figs. 1(b) and 1(c), the $\nu$ and $\bar \lambda$ diagrams are reproduced from Ref. \cite{AWNPA}, with the lines 
of Fig. 1(a) plotted on them. We see that the present line remains within the valley corresponding to the arc of regularity 
for most of the way from the SU(3) vertex towards the critical line, turning upwards a little before reaching 
the critical line. 

It should be noted that the present study is greatly facilitated by the fact that the position of the arc of regularity 
appears to be practically independent of the number of bosons, as already remarked in Refs. \cite{AWPRL,AWNPA} 
and corroborated in Ref. \cite{arcPRL}. 

In summary, we have achieved by now two goals.

1) To prove analytically the existence of a line in the parameter space of the IBA, along which the Hamiltonian  
approximately commutes with the SU(3) generators in the large $N$ limit. 

2) To prove that this line closely follows the Alhassid--Whelan arc of regularity in the region between the SU(3) vertex 
and the critical line of first order shape/phase transition. 

\subsection{Matrix elements}

In going from Eq. (\ref{interm4}) to Eq. (\ref{inter5}), we replaced the quadrupole operator $\hat Q^{(2)}_{SU(3)}$ 
by the intrinsic quadrupole moment. To justify this, 
we consider here the matrix elements of $[\hat H_1, \hat Q^{(2)}_{SU(3),\nu}]$ and  $[\hat H_2, \hat Q^{(2)}_{SU(3),\nu}]$
within the ground state band. We will examine the conditions under which these matrix elements lead to a vanishing result. 
In this calculation the intrinsic quadrupole moment appears naturally when calculating the matrix elements of the 
commutators of the relevant parts of the Hamiltonian with the quadrupole operator,
and not as a result of an approximation, as in the previous subsection. 

Using the standard formalism for treating matrix elements of the tensor product of two tensor operators 
and the single-boson matrix elements given in Table 1, one finds for the first term of the Hamiltonian 
\begin{eqnarray}\label{meH_1}
\langle [N],(2N,0),\tilde\chi=0,L||[\hat H_1, \hat Q^{(2)}_{SU(3)}]|| [N],(2N,0),\tilde\chi=0,L \rangle \nonumber\\
= -{2\over 3\sqrt{7}} c \eta  N  R_1 \sqrt{2L+1}, 
\end{eqnarray}
where
\begin{equation}\label{R1}
R_1 =  {(2N-L)(2N+L+1)\over (2N-1) (2N)} {(2L-3)(2L+5)\over (2L-1)(2L+3)}, 
\end{equation}
and $\tilde \chi$ is the Vergados quantum number \cite{Vergados}, not to be confused with the parameter $\chi$
of the Hamiltonian (\ref{HAW}). 
It is worth remarking that the terms $d^\dagger s$ and $s^\dagger \tilde d$ give equal contributions 
of the same sign to the final result, despite the fact that they appear in Eq. (\ref{term1}) with opposite signs. 

Following the procedure described in Appendix 3, one finds for the second term of the Hamiltonian, 
including in the large $N$ limit only the terms appearing in the second line of Eq. (\ref{interm4}) 
after replacing $\hat Q^{(2)}_\chi$ by $\hat Q^{(2)}_{SU(3)}$
\begin{eqnarray}\label{meH_2}
\langle [N],(2N,0),\tilde\chi=0,L||[\hat H_2, \hat Q^{(2)}_{SU(3)}]|| [N],(2N,0),\tilde\chi=0,L \rangle \nonumber\\
= {1\over 7\sqrt{2}} c (1-\eta)
 \left( \chi+ {\sqrt{7}\over 2} \right) q_0 R_2 \sqrt{2L+1},
\end{eqnarray}  
where
\begin{eqnarray}\label{R2}
R_2= \left[ {4\over 3} {(2N-L)(2N+L+1)\over (2N-1)(2N)} {(2L-3)^2 (2L+5)^2 \over (2L-1)^2 (2L+3)^2} \right. \nonumber\\
+{(2N-L-2)\sqrt{(2N-L)(2N+L+3)} \over (2N-1)(2N)} {\sqrt{(2L-1)(2L)(2L+2)(2L+4)(2L+6)(2L+7)} \over (2L+1)^2 (2L+3)^2} \nonumber \\
+{(2N+L-1)\sqrt{(2N-L-2)(2N+L+1)} \over (2N-1)(2N)} {\sqrt{(2L-5)(2L-4)(2L-2)(2L)(2L+2)(2L+3)} \over (2L+1) (2L-1)^2} \nonumber \\
+{(2N+L+1)\sqrt{(2N-L)(2N+L+3)} \over (2N-1)(2N)} {(2L-1)(2L)(2L+4)\over (2L+1) (2L+3)^2} \nonumber \\
\left. +{(2N-L)\sqrt{(2N-L+2)(2N+L+1)} \over (2N-1)(2N)} {(2L-2)(2L+2)(2L+3)\over (2L+1) (2L-1)^2} \right]. \nonumber\\
\end{eqnarray}
The only approximation made in the derivation of this equation in Appendix 3 has been the replacement of the reduced matrix 
elements of the quadrupole operator by their values in the contraction limit, which contain the intrinsic quadrupole moment $q_0$, 
according to Eq. (\ref{Vass}). Again the terms $d^\dagger s$ and $s^\dagger \tilde d$ give equal contributions 
of the same sign to the final result, despite the fact that they appear in Eq. (\ref{interm4}) with opposite signs. 

One can simplify Eqs. (\ref{R1}) and (\ref{R2}) by making the following approximations

\noindent 
1) Ratios of terms containing $2N$ can be replaced by unity, since we work in the large $N$ limit.

\noindent 
2) Ratios of terms containing $2L$ can be replaced by unity, if $L$ is not too small. We will show below numerically 
that this approximation is an accurate one.  

Using these approximations we obtain $R_1=1$ and $R_2=16/3$.  
Replacing then the intrinsic quadrupole moment $q_0$ by its value ($N\sqrt{2}$) from Eq. (\ref{prolate}), we see that the two matrix 
elements vanish in the large $N$ limit if the condition 
\begin{equation}\label{gsb_AW} 
\chi(\eta) = {\sqrt{7}\over 8} {\eta \over (1-\eta)} - {\sqrt{7}\over 2}
\end{equation}
is fulfilled. 

We remark that this condition is very similar to Eq. (\ref{line_eta}), since $\sqrt{7}/8= 0.3307$ 
appears in the former, while $1/(2\sqrt{2})=0.3536$ appears in the latter. This can be visualized in Fig.~2,
in which these two conditions are compared to the original expression for the arc, Eq. (\ref{arc}). 

It should be noticed that Eq. (\ref{gsb_AW}) has been derived in the special case of considering matrix elements within 
the ground state band only. This is similar to the condition discussed earlier in the context of Eq. (\ref{interm}), 
where the $Q^{(2)}_{SU(3)}$ operators simplify when the ground state band becomes energetically isolated.

Concerning the accuracy of the approximations made above, one can check numerically the ratio $R_2/R_1$, for which the value $16/3$
has been used. For $N=250$ (the boson number used in Ref. \cite{arcPRL}) one can easily see that the exact values of $R_2/R_1$ 
deviate by less than 1\% from the approximate value $16/3$ for angular momenta between 8 and 88, while they deviate 
by less than 5\% for angular momenta between 4 and 176. Therefore the approximation is accurate for a large fraction of the spectrum,
the lower-lying one. In the present case of $N=250$ the ground state band extends up to $L=500$, thus the approximation 
is good (deviations less than 5\%) for the lowest 1/3 of the spectrum. 
 
It should be clarified that in both subsections 3.1 and 3.2 the intrinsic quadrupole moment is finally replaced by its value, given 
by Eq. (\ref{prolate}). The main difference, though, is that in subsection 3.1 the intrinsic quadrupole moment appears 
by approximating the quadrupole operator by the intrinsic quadrupole moment, by looking at the SU(3) Casimir operator, as explained 
in detail below Eq. (\ref{prolate}), while in subsection 3.2 the intrinsic quadrupole moment arises through the proper detailed 
calculation of the matrix elements of the commutators of the relevant parts of the Hamiltonian with the quadrupole operator
within the ground state band. 
Therefore subsection 3.2 serves as a detailed justification of the approximation made in subsection 3.1~.
This detailed justification has been carried out within the ground state band, which suffices in the present case, 
since we work in the contraction limit, in which the ground state band gets isolated from the rest of the bands.   

It should be further noted that the method used in the present subsection, i.e., the detailed calculation of matrix elements,
leading to Eq. (\ref{gsb_AW}), allows for an estimation of the accuracy of the approximations used within the ground state band
for each value of $L$, as discussed above for the $N=250$ case. In contrast, the method used in subsection 3.1, 
leading to Eq. (\ref{line_eta}), uses an approximation (the replacement of the quadrupole operator by the intrinsic quadrupole moment), 
which cannot be tested separately for each state. The same holds for the work of Alhassid and Whelan \cite{AWPRL,AWNPA}, 
leading to the regular region approximated \cite{CJPRE} by Eq. (\ref{arc}). Indeed, in Refs. \cite{AWPRL,AWNPA}, the statistical properties 
of the spectrum as a whole are considered, for $N=25$. This implies that numerical studies for large boson numbers ($N=250$, for example), 
using statistical tools, should be undertaken, considering the spectrum as a whole, as well as by parts (lowest 1/3, middle 1/3, highest 1/3,
for example). Studies of this kind should further clarify the relation between the present results and the Alhassid--Whelan arc of regularity.

\section{The $\overline{\rm \bf SU(3)}$ symmetry} 

The question is now raised about what happens in the triangle formed by U(5), O(6), and $\overline{\rm SU(3)}$ \cite{IA}, 
the algebra containing the quadrupole operators with $\chi=+\sqrt{7}/2$, which is known to correspond to oblate nuclei, 
while SU(3) is related to prolate nuclei.  

It turns out that the relevant calculation follows the same steps, with two notable differences:

1) $-\sqrt{7}/2$ is replaced by $+\sqrt{7}/2$ everywhere.

2) The intrinsic quadrupole moment changes sign (see Appendix 2), in agreement to the well known fact that the intrinsic 
quadrupole moment has positive values for prolate nuclei and negative values for oblate nuclei \cite{triangle}.  

As a result of these two changes, Eq. (\ref{line_eta}) takes in the large $N$ limit the form
\begin{equation}\label{linem}
\chi(\eta) = -{1\over 2\sqrt{2}} {\eta \over (1-\eta)} + {\sqrt{7}\over 2},
\end{equation}
It becomes then clear that for a given $\eta$ in this case  $\chi$ acquires  the opposite value 
from the one it gets within the U(5)-O(6)-SU(3) triangle. This is in agreement with the well known fact 
that properties within the U(5)-O(6)-$\overline{\rm SU(3)}$ triangle are mirror images of the properties 
appearing within the U(5)-O(6)-SU(3) triangle \cite{IA}. 

\section{The O(6) symmetry} 

The successful determination of a line in the symmetry triangle characterized by approximate SU(3) 
symmetry, raises the question if a similar line related to the O(6) symmetry can be determined. 
The O(6) algebra \cite{IA} is generated by the angular momentum operators of Eq. (\ref{L}) and the operators 
$(d^\dagger \tilde d)^{(3)}_\xi$, forming together the O(5) subalgebra, plus the quadrupole operators 
\begin{equation} 
\hat Q^{(2)}_{O(6),\xi} = (s^\dagger \tilde d + d^\dagger s)^{(2)}_\xi.
\end{equation}
Following the same procedure as in Section 3, we are going to examine the conditions under which 
the Hamiltonian commutes with the generators $\hat Q^{(2)}_{O(6),\xi}$. The needed commutation relations are listed 
in Appendix~1. 

The first term of the Hamiltonian gives 
\begin{equation}\label{term1b}
[\hat H_1, \hat Q^{(2)}_{O(6),\nu}] = c \eta [\hat n_d, \hat Q^{(2)}_{O(6),\nu}] = c \eta [ (d^\dagger s -s^\dagger \tilde d)^{(2)}_\nu,
\end{equation}
which is similar to Eq. (\ref{term1}). 

Using 
\begin{equation}
\hat Q ^{(2)}_{\chi,\xi} = \hat Q ^{(2)}_{O(6),\xi} + \chi (d^\dagger \tilde d)^{(2)}_\xi ,  
\end{equation}
in the second term of the Hamiltonian, one gets the intermediate result
$$
[\hat Q^{(2)}_\chi \cdot \hat Q^{(2)}_\chi, \hat Q^{(2)}_{O(6),\nu}] = 
\sum_\xi (-1)^\xi \left\{ [\hat Q^{(2)}_{O(6),\xi}, \hat Q^{(2)}_{O(6),\nu}] \hat Q^{(2)}_{x, -\xi} + 
\hat Q^{(2)}_{x, \xi} [\hat Q^{(2)}_{O(6),-\xi}, \hat Q^{(2)}_{O(6),\nu}] \right.$$
\begin{equation}\label{interm2} \left.
+  \chi  \left\{ [ (d^\dagger \tilde d)^{(2)}_\xi , \hat Q^{(2)}_{O(6), \nu} ] \hat Q^{(2)}_{x, -\xi}
+ \hat Q^{(2)}_{x, \xi} [ (d^\dagger \tilde d)^{(2)}_{-\xi} , \hat Q^{(2)}_{O(6), \nu} ].
\right\} \right\} 
\end{equation}
Again this expression can be simplified in the large $N$ limit, since in this limit the contraction 
of O(6) to R$^5$[SO(5)] \cite{MtV84,Elliott171} takes place (see Appendix 4). This means that the commutators in the first two terms will vanish. 
Since 
\begin{equation}
[ \hat Q^{(2)}_{O(6),\xi},  \hat Q^{(2)}_{O(6),\nu}] = 2 (2 \xi 2 \nu | 1 \xi+\nu) (d^\dagger \tilde d)^{(1)}_{\xi+\nu}  
+ 2 (2 \xi 2 \nu | 3 \xi+\nu) (d^\dagger \tilde d)^{(3)}_{\xi+\nu},   
\end{equation}
the vanishing of this commutator implies that terms of the form $(d^\dagger \tilde d)^{(k)}$ can be ignored.

Eq. (\ref{interm2}) can be rewritten, without using any approximations yet, as 
$$
[\hat Q^{(2)}_\chi \cdot \hat Q^{(2)}_\chi, \hat Q^{(2)}_{O(6),\nu}] = 
\sum_{k=1,3} 2 \sqrt{2k+1\over 5}  [ ((d^\dagger \tilde d)^{(k)} \hat Q^{(2)}_{x})^{(2)} _\nu -
(\hat Q^{(2)}_{x}  (d^\dagger \tilde d)^{(k)} )^{(2)}_\nu ] $$
\begin{equation} \label{basic} 
+ \chi  [((d^\dagger s- s^\dagger \tilde d)^{(2)} \hat Q^{(2)}_\chi )^{(2)}_\nu 
+ ( \hat Q^{(2)}_\chi   (d^\dagger s -s^\dagger \tilde d)^{(2)} )^{(2)}_\nu ] . 
\end{equation} 
In the large $N$ limit the terms containing $(d^\dagger \tilde d)^{(k)}$ can be omitted. 
In addition, in this limit $\hat Q^{(2)}_{O(6)}$ can be replaced by the intrinsic quadrupole moment (a scalar),
which is $N$ in the present case (see Appendix 4). 
Then in the large $N$ limit one is left with 
\begin{equation}
[\hat Q^{(2)}_\chi \cdot \hat Q^{(2)}_\chi, \hat Q^{(2)}_{O(6),\nu}] = 2  N  \chi 
[ (d^\dagger s -s^\dagger \tilde d)^{(2)}_\nu.
\end{equation}
The commutator for the second part of the Hamiltonian in the large $N$ limit then reads
\begin{equation}\label{term2b}
[\hat H_2, \hat Q^{(2)}_{O(6),\nu}]= - 2 c (1-\eta) \chi [ (d^\dagger s -s^\dagger \tilde d)^{(2)}_\nu.
\end{equation}
In order to get a vanishing commutator, the coefficients of $ (d^\dagger s -s^\dagger \tilde d)^{(2)}_\nu $ in Eqs. (\ref{term1b}) 
and (\ref{term2b}) should cancel, leading in the large $N$ limit to the condition 
\begin{equation}\label{line2}
\chi(\eta)   = {\eta \over 2(1-\eta)}. 
\end{equation}
However, this does not suffice yet to guarantee the existence of a line corresponding to the O(6) symmetry.
One has also to consider the commutators of the Hamiltonian with the generators of O(5), to be considered 
in the next Section. 

\section{The O(5) symmetry} 

As it has already been mentioned, the O(5) algebra is generated by the angular momentum operators of Eq. (\ref{L}) and the operators 
$(d^\dagger \tilde d)^{(3)}_\xi$. We should examine the conditions under which the commutator of the Hamiltonian 
with $(d^\dagger \tilde d)^{(3)}_\xi$ vanishes.

In this case the first term of the Hamiltonian makes no contribution, since $\hat n_d$ is known to be an O(5) scalar \cite{Talmi}. 
Taking advantage of the fact that
 $\hat Q^{(2)}_{O(6)} \cdot \hat Q^{(2)}_{O(6)}$ is also an O(6) scalar \cite{Talmi}
in order to simplify the calculation, one obtains 
$$
[\hat H, (d^\dagger \tilde d)^{(3)}_\nu ] = {15\over 7} \chi [ ( (d^\dagger \tilde d)^{(2)} (s^\dagger \tilde d + d^\dagger s)^{(2)} )^{(3)}_\nu 
- ((s^\dagger \tilde d + d^\dagger s)^{(2)} (d^\dagger \tilde d)^{(2)})^{(3)}_\nu ] $$
$$
- {3 \over 7} \sqrt{10} \chi [ ( (d^\dagger \tilde d)^{(4)} (s^\dagger \tilde d + d^\dagger s)^{(2)} )^{(3)}_\nu 
- ((s^\dagger \tilde d + d^\dagger s)^{(2)} (d^\dagger \tilde d)^{(4)})^{(3)}_\nu ] $$
\begin{equation}
- {3 \over 7} \sqrt{10} \chi^2  [ ( (d^\dagger \tilde d)^{(4)} (d^\dagger \tilde d)^{(2)} )^{(3)}_\nu 
- ((d^\dagger \tilde d)^{(2)} (d^\dagger \tilde d)^{(4)})^{(3)}_\nu ].
\end{equation}
In this case no simplification due to contractions can be made. In order to get a vanishing commutator one needs 
to put $\chi=0$, thus being confined on the U(5)-O(6) side of the triangle. This finding is in agreement with the known 
existence of an O(5) subalgebra along the U(5)-O(6) side of the triangle \cite{Talmi}. 

Going back to the question of the existence of a line related to the O(6) symmetry, we see that in Eq. (\ref{line2})
one has to put $\chi=0$, as required by the O(5) subalgebra. Then one ends up with $\eta=0$, which represents the O(6) vertex alone.
Thus no line related to the O(6) symmetry exists within the triangle.  

\section{A different parametrization} 

In recent years, an IBA Hamiltonian in common use reads 
\cite{Werner,Zamfir66}
\begin{equation}\label{HIBA}
\hat H(\zeta,\chi) = a \left[ (1-\zeta) \hat n_d -{\zeta\over 4 N_B}
\hat Q^{(2)}_\chi \cdot \hat Q^{(2)}_\chi\right],
\end{equation}
where $c$ is a scaling factor, while the rest of the symbols have the same meaning 
as in Eq. (\ref{HAW}).
The parameter $\zeta$ ranges from 0 to 1, while the parameter 
$\chi$ ranges from 0 to $-\sqrt{7}/2$, as above. 
The parametrizations of the Hamiltonians (\ref{HAW}) and (\ref{HIBA}) are related by 
\cite{AWexp} 
\begin {equation}\label{eta}
\eta={4(\zeta-1)\over 3\zeta-4}, \qquad c ={1\over 4} a (4-3\zeta).  
\end{equation}
From these relations and the results reported at the end of subsection 3.1 
we find that for large $N$ the critical line at $\chi=0$ corresponds to 
$\zeta_{crit}=0.5$, while at $\chi=-\sqrt{7}/2$ 
it corresponds to $\zeta_{crit}=8/17=0.471$.

Eq. (\ref{line_eta}), valid in the large $N$ limit, in this parametrization reads
\begin{equation}\label{line}
\chi(\zeta) = \sqrt{2} \left ({1\over \zeta}-1\right) - {\sqrt{7}\over 2},  
\end{equation}
while Eq. (\ref{gsb_AW}), again valid in the large $N$ limit, reads 
\begin{equation}\label{gsb}
\chi(\zeta) = {\sqrt{7}\over 2} \left ({1\over \zeta}-1\right) - {\sqrt{7}\over 2} =  {\sqrt{7}\over 2} \left ({1\over \zeta}-2\right). 
\end{equation}

\section{Discussion} 

In the present work we find the first analytical evidence for the existence of an approximate symmetry 
inside the symmetry triangle of the IBA. The SU(3) symmetry found extends from the SU(3) vertex 
until close to the critical line separating the spherical and prolate deformed shapes/phases, following 
the Alhassid--Whelan arc of regular behavior amidst chaotic regions. Thus it also points to 
an underlying SU(3) symmetry as an explanation for the existence of this arc of regularity.
The present line of SU(3) symmetry has been determined in the limit of large boson numbers, 
taking advantage of the contraction of SU(3) to [R$^5$]SO(3), the algebra of the rigid rotator, in this limit. 
The proof is valid for the lower part of the spectrum, since SU(3) irreps with $\lambda$ much higher 
than the angular momentum $L$ have been used in the contraction procedure. 
The contraction of O(6) to [R$^5$]SO(5), the algebra of the $\gamma$-unstable rotator,
 has also been worked out, leading to a similar line,
which however is shrinked to a point when the O(5) symmetry is imposed. 

The approximate symmetry determined here bears some similarity to the concept of 
quasidynamical symmetries (QDS) \cite{Rowe1225,Rowe2325,Rowe745,Rowe756,Rowe759}, in which 
several features of a symmetry persist far outside its expected region of applicability.
A difference is that in the present approach an analytical proof of the existence of the symmetry 
is offered, while QDS have been so far located numerically. 

A concept related to the present work is that of partial dynamical symmetries (PDS) \cite{AL25,Lev77,LVI89,Lev2010},
situations in which part of the states preserve all the dynamical symmetry (PDS type I), 
or all the states preserve part of the dynamical symmetry (PDS type II), or part of the states 
preserve part of the dynamical symmetry (PDS type III). The present approach can be seen as an analog 
of a PDS of type I, since part of the states (the low lying ones) preserve all the dynamical symmetry.  

The method used in the present work is of wider applicability. It can be used for algebraic 
Hamiltonians known to possess symmetries due to existing subalgebras, materialized 
for specific parameter values. The present method allows the extension of the region of the 
parameter space of the Hamiltonian in which the specific symmetry appears, in cases in which 
the relevant subalgebra contracts to a different algebra in some limiting case (the large boson 
number limit in the present case). 

The application  of the present techniques to IBA-2 \cite{IA}, in which distinction between protons and neutrons 
is made, to the sdg-IBA model \cite{IA,Kota}, in which the $L=4$ boson is taken into account, to the spdf-IBA model \cite{EINPA,Kusn1,Kusn2}, 
in which negative parity bosons with $L=1$, 3 are included, and to the schematic Hamiltonian of Ref. \cite{AL25,Lev179,WAL}, 
which exhibits a PDS, are interesting tasks. Shape/phase transitions have recently been studied in both the parameter space of IBA-2,
which is a tetrahedron  \cite{AriasPRL,CIPRL,CIAP}, and the parameter space of sdg-IBA \cite{PVI}, which is a prism. 

\section*{Acknowledgements} 

Earlier numerical calculations on the arc, carried out with E. A. McCutchan and reported in Ref. \cite{arcPRL}, 
using the IBAR code \cite{IBAR1,IBAR2} for IBA calculations involving large boson numbers developed by R. J. Casperson, influenced 
the present development. Work supported by U.S.~DOE Grant No.~DE-FG02-91ER-40609.

\section*{Appendix 1. Commutation relations.}

We list here the commutation relations needed for obtaining the results of Sections 3--6. 
They are obtained through standard angular momentum coupling techniques \cite{Edmonds}. 
\begin{equation}
[\hat n_d, (d^\dagger s+s^\dagger \tilde d)^{(2)}_\xi] = (d^\dagger s - s^\dagger \tilde d)^{(2)}_\xi,
\end{equation}
\begin{equation}
[(d^\dagger \tilde d)^{(2)}_\xi, (d^\dagger s+s^\dagger \tilde d)^{(2)}_\nu] = (2 \xi 2 \nu | 2 \xi+\nu) 
(d^\dagger s - s^\dagger \tilde d)^{(2)}_{\xi+\nu},
\end{equation}
\begin{equation}
[\hat n_d, (d^\dagger \tilde d)^{(k)}_\xi]=0, \quad k=0,1,2,3,4, 
\end{equation}
\begin{equation}
[(d^\dagger \tilde d)^{(2)}_\xi, (d^\dagger \tilde d)^{(2)}_\nu]= -10 \sum_{k=1,3} (2 \xi 2 \nu | k \xi+\nu) 
\left\{ \matrix{2 & 2 & k \cr 2 & 2 & 2 \cr} \right\} (d^\dagger \tilde d)^{(k)}_{\xi +\nu},
\end{equation}
\begin{equation}
[(d^\dagger s+s^\dagger \tilde d)^{(2)}_\xi, (d^\dagger s+s^\dagger \tilde d)^{(2)}_\nu] = 2 \sum_{k=1,3} (2 \xi 2 \nu | k \xi+\nu)
(d^\dagger \tilde d)^{(k)}_{\xi +\nu},
\end{equation}
\begin{equation}
[(d^\dagger s+s^\dagger \tilde d)^{(2)}_\xi, (d^\dagger \tilde d)^{(1)}_\nu] = -\sqrt{3\over 5} (2 \xi 1 \nu | 2 \xi+\nu) 
(d^\dagger s+s^\dagger \tilde d)^{(2)}_{\xi+\nu},
\end{equation}
\begin{equation}
[(d^\dagger s+s^\dagger \tilde d)^{(2)}_\xi, (d^\dagger \tilde d)^{(3)}_\nu] = -\sqrt{7\over 5} (2 \xi 3 \nu | 2 \xi+\nu) 
(d^\dagger s+s^\dagger \tilde d)^{(2)}_{\xi+\nu},
\end{equation}
\begin{equation}
[(d^\dagger \tilde d)^{(1)}_\xi, (d^\dagger \tilde d)^{(1)}_\nu]=-{1\over \sqrt{5}} (1 \xi 1 \nu | 1 \xi+\nu) (d^\dagger \tilde d)^{(1)}_{\xi +\nu},
\end{equation}
\begin{equation}
[(d^\dagger \tilde d)^{(2)}_\xi, (d^\dagger \tilde d)^{(1)}_\nu]= 2 \sqrt{15}  (2 \xi 1 \nu | 2 \xi+\nu) 
\left\{ \matrix{2 & 2 & 1 \cr 2 & 2 & 2 \cr} \right\} (d^\dagger \tilde d)^{(2)}_{\xi +\nu},
\end{equation}
\begin{equation}
[(d^\dagger \tilde d)^{(2)}_\xi, (d^\dagger \tilde d)^{(3)}_\nu]= 2 \sqrt{35} \sum_{k=2,4} (2 \xi 3 \nu | k \xi+\nu) 
\left\{ \matrix{2 & 3 & k \cr 2 & 2 & 2 \cr} \right\} (d^\dagger \tilde d)^{(k)}_{\xi +\nu}.
\end{equation}

\section*{Appendix 2. The SU(3)$\to$ [R$^5]$SO(3) contraction}

The SU(3)$\to$ [R$^5]$SO(3) contraction has been studied in Refs. \cite{LeBlanc,RowePPNP}.
It is a procedure in which the full SU(3) algebra, consisting of 8 noncommuting generators, is shrinked
into an SO(3) algebra (consisting of 3 noncommuting generators), accompanied by 5 mutually commuting 
operators (the quadrupole operators). 
This simplification occurs in the limit of large boson number in which, in SU(3), 
all intrinsic excitations rise in energy, isolating the ground state band so that SU(3) goes over, 
approximately, into a simple rigid rotator.
The resulting  algebraic structure is, indeed, known \cite{Ui} to be the algebra 
of the rigid rotator. 
The scaling of Ref. \cite{IA} is used here. (The quadrupole operator in Refs. \cite{LeBlanc,RowePPNP}
is $2\sqrt{2}$ times the quadrupole operator of Ref. \cite{IA}.) 

The SU(3) commutation relations read
\begin{equation}
[\hat L_\xi, \hat L_\nu]= -\sqrt{2} (1 \xi 1 \nu | 1 \xi+\nu) \hat L_{\xi+\nu},
\end{equation}
\begin{equation}
[\hat L_\xi, \hat Q^{(2)}_{SU(3),\nu}] = -\sqrt{6} ( 1\xi 2 \nu | 2 \xi+\nu) \hat Q^{(2)}_{SU(3),\xi+\nu},  
\end{equation}
\begin{equation}
[\hat Q^{(2)}_{SU(3),\xi}, \hat Q^{(2)}_{SU(3),\nu}] = {3\over 4} \sqrt{5 \over 2} (2 \xi 2 \nu | 1 \xi+\nu) \hat L_{\xi+\nu}.
\end{equation}
The second order Casimir operator is 
\begin{equation}\label{SU3C}
\hat C_2 [SU(3)]= {2\over 3} \left[ 2 \hat Q_{SU(3)}^{(2)} \cdot \hat Q_{SU(3)}^{(2)} +{3\over 4} \hat L \cdot \hat L\right], 
\end{equation}
while its eigenvalues in the Elliott basis, $(\lambda, \mu)$, are 
\begin{equation}\label{SU3c}
C_2(\lambda,\mu)= {2\over 3} (\lambda^2 + \mu^2 + \lambda \mu + 3 \lambda + 3 \mu).
\end{equation}
If we consider SU(3) irreducible representations (irreps) with  large values of $C_2(\lambda,\mu)$,
that is for large boson numbers,
we can rescale the quadrupole operator as 
\begin{equation}
\hat q^{(2)}_{SU(3),\xi}= { \hat Q^{(2)}_{SU(3),\xi} \over \sqrt{C_2(\lambda,\mu) } }.
\end{equation}
The first two commutation relations remain unchanged by the rescaling, while the last one becomes 
\begin{equation}\label{qqSU3}
[\hat q^{(2)}_{SU(3),\xi}, \hat q^{(2)}_{SU(3),\nu}] = {3\over 4} \sqrt{5 \over 2} (2 \xi 2 \nu | 1 \xi+\nu) {\hat L_{\xi+\nu} \over
C_2(\lambda,\mu)}.
\end{equation}
Then in the limit of large values of $C_2(\lambda,\mu)$ one gets 
\begin{equation}\label{qqSU3b}
[\hat q^{(2)}_{SU(3),\xi}, \hat q^{(2)}_{SU(3),\nu}] = 0.
\end{equation}
This result, which is obtained for large boson number, is called the contraction of SU(3) to [R$^5]$SO(3), where [R$^5]$SO(3) is the algebra 
of the rigid rotator\cite{Ui}, generated by the angular momentum operators of SO(3) and the five commuting 
operators $\hat q^{(2)}_{SU(3),\xi}$, $\xi=-2,-1,0,1,2$.

An immediate consequence of Eqs. (\ref{qqSU3}) and (\ref{qqSU3b}) is that, in the contraction limit, terms proportional to 
the angular momentum $\hat L$ can be ignored. In the IBA framework, in which $\hat L$ is proportional to  $(d^\dagger \tilde d)^{1}$,
as seen in Eq. (\ref{L}), this implies that $(d^\dagger \tilde d)^{1}$ terms can be ignored. 

In the limit of large values of $C_2(\lambda,\mu)$ and $\lambda \geq \mu$ the intrinsic quadrupole 
moments become \cite{RowePPNP,Vassanji}
\begin{equation}\label{q0q2}
q_0= {1\over 2\sqrt{2}} (2\lambda + \mu +3), \qquad q_2= {1\over 4} \sqrt{3 (\mu-K)(\mu+K+2)},
\end{equation}
where $K$ is the eigenvalue of the angular momentum projection on the body-fixed $z$-axis, 
for which $K\leq L$ is valid, as one can see from the algorithm of the SU(3)$\supset$SO(3) reduction \cite{IA}.   
(Remember at this point that the quadrupole operator used in Refs. \cite{LeBlanc,RowePPNP,Vassanji} is $2\sqrt{2}$ times the quadrupole operator 
used in the present work.)
For states with $\lambda >>L$ (therefore also $\lambda>>K$) and $\lambda >>\mu$ one then obtains \cite{LeBlanc}
\begin{equation}
q_0={\lambda \over \sqrt{2}},
\end{equation}
while $q_2$ becomes negligible. 
Since the ground state band belongs to the $(2N,0)$ irreducible representation (irrep) of SU(3), 
while other low-lying bands belong to irreps $(2N-4i-6j,2i)$, $i=0$,1,2,\dots, $j=0$,1,2,\dots 
with relatively low $i$, $j$, the contraction occurs in the large $N$ limit. 
Thus in the case of interest the intrinsic quadrupole moment becomes 
\begin{equation}\label{prolate}
q_0= N \sqrt{2}.  
\end{equation}
An equivalent statement is that one can approximately replace the operator $\hat Q_{SU(3)}^{(2)}$ 
by the scalar $\lambda/\sqrt{2}$, as one can see from Eqs. (\ref{SU3C}) and (\ref{SU3c}), since 
the terms containing $\hat L$ and $\mu$ are negligible in this limit, having as a consequence that 
only the first term in the rhs of these equations survives. A formal justification 
for this replacement is given in subsection 3.2 and Appendix 3, where matrix elements of the commutators 
of the relevant parts of the Hamiltonian with the quadrupole operator 
are properly considered, resulting in the appearance of the intrinsic quadrupole moment. 
 
It should be noticed that the above results have been obtained in irreps with $\lambda >> L$, 
thus they regard the low lying part of the spectrum. 

In SU(3) the irreps are built out of the (2,0) irrep, while in the case of $\overline{\rm SU(3)}$ 
the irreps are built out of the (0,2) irrep \cite{IA}. As a result, in the $\overline{\rm SU(3)}$
framework one is interested in states with large values of $C_2(\lambda,\mu)$ and $\lambda < \mu$, 
in which the intrinsic quadrupole 
moments become \cite{RowePPNP,Vassanji}
\begin{equation}
q_0= -{1\over 2\sqrt{2}} (\lambda + 2\mu +3), \qquad q_2= -{1\over 4} \sqrt{3 (\lambda-K)(\lambda+K+2)}.
\end{equation}  
For states with $\mu >>L$ and $\mu >>\lambda$ one then obtains 
\begin{equation}
q_0=-{\mu \over \sqrt{2}}, 
\end{equation}
while $q_2$ becomes negligible. 
Since the ground state band belongs to the $(0,2N)$ irrep of $\overline{\rm SU(3)}$, 
while other low-lying bands belong to irreps $(2i,2N-4i-6j)$, $i=0$,1,2,\dots, $j=0$,1,2,\dots 
with relatively low $i$, $j$, the contraction does occur in the large $N$ limit, the intrinsic quadrupole moment becoming 
\begin{equation}\label{oblate}
q_0= -N \sqrt{2}.  
\end{equation}
Since SU(3) is associated to prolate shapes, while $\overline{\rm SU(3)}$ is related to oblate shapes, 
the signs in Eqs. (\ref{prolate}) and (\ref{oblate}) are consistent with the fact that intrinsic quadrupole moments 
are known to be positive for prolate nuclei and negative for oblate nuclei \cite{triangle}. 

\section*{Appendix 3. Matrix elements} 

In order to show how Eq. (\ref{meH_2}) is derived, we consider in detail the matrix element 
\begin{eqnarray}\label{dsQ}
\langle [N], (2N,0), \tilde\chi=0, L || ((d^\dagger s)^{(2)} Q_{SU(3)}^{(2)})^{(2)} ||  [N], (2N,0), \tilde\chi=0, L \rangle \nonumber \\ 
= \sqrt{5} \sum_{L''} \left\{ \matrix{2 & 2 & 2 \cr L & L & L'' \cr} \right\}
\langle [N], (2N,0), \tilde\chi=0, L || (d^\dagger s)^{(2)} ||  [N], (2N,0), \tilde\chi=0, L'' \rangle \nonumber \\ 
\langle [N], (2N,0), \tilde\chi=0, L'' || Q_{SU(3)}^{(2)} ||  [N], (2N,0), \tilde\chi=0, L \rangle, 
\end{eqnarray}
where the standard formalism concerning the matrix elements of the tensor product of two tensor operators \cite{Edmonds} has been used. 
By $\tilde \chi$ we denote the Vergados quantum number \cite{Vergados} [not to be confused with the parameter $\chi$ 
of the Hamiltonian (\ref{HAW})], which corresponds to an orthogonal basis, 
while the Elliott quantum number $K$ \cite{ElliottII}, coinciding with the angular momentum projection on the body-fixed 
$z$-axis used in Appendix 2, corresponds to a non-orthogonal basis.   

This expression can be simplified since in the contraction limit the matrix elements of the quadrupole operator become 
\cite{Vassanji}
\begin{equation}\label{Vass}
\langle [N], (2N,0), \tilde\chi=0, L' || Q_{SU(3)}^{(2)} ||  [N], (2N,0), \tilde\chi=0, L \rangle
=  \sqrt{2L+1} (L 0 2 0 | L' 0) q_0, 
\end{equation}
where $q_0$ is the intrinsic quadrupole moment of Eq. (\ref{q0q2}). This can be seen from the Elliott matrix element 
\cite {ElliottII} of the quadrupole operator.

Using Eq. (\ref{Vass}) in Eq. (\ref{dsQ}) one obtains in the contraction limit 
\begin{eqnarray}
\langle [N], (2N,0), \tilde\chi=0, L || ((d^\dagger s)^{(2)} Q_{SU(3)}^{(2)})^{(2)} ||  [N], (2N,0), \tilde\chi=0, L \rangle 
= \sqrt{5} q_0 \sqrt{2L+1}  \nonumber \\
\sum_{L''=L,L\pm 2} \left\{ \matrix{2 & 2 & 2 \cr L & L & L'' \cr} \right\} (L 0 2 0 | L'' 0)
\langle [N], (2N,0), \tilde\chi=0, L || (d^\dagger s)^{(2)} ||  [N], (2N,0), \tilde\chi=0, L'' \rangle. 
\end{eqnarray}
The matrix elements appearing in the last equation can be written in terms of matrix elements of single boson 
operators 
\begin{eqnarray}
\langle [N], (2N,0), \tilde\chi=0, L || (d^\dagger s)^{(2)} ||  [N], (2N,0), \tilde\chi=0, L'' \rangle =
\sqrt{5} \left\{ \matrix{2 & 2 & 2 \cr L'' & L & L'' \cr} \right\} \nonumber \\ 
\langle [N], (2N,0), \tilde\chi=0, L || d^\dagger ||  [N-1], (2N-2,0), \tilde\chi=0, L'' \rangle \nonumber \\
\langle [N-1], (2N-2,0), \tilde\chi=0, L'' || s ||  [N], (2N,0), \tilde\chi=0, L'' \rangle.
\end{eqnarray}
The matrix elements of single boson operators needed here and in subsequent calculations are listed in Table~1. 
Some of them are given in Ref. \cite{IA}, while the rest have been calculated following the method of Ref. \cite{Rosensteel},
using the triple-barred SU(3) reduced matrix elements given there and the SU(3)$\supset$SO(3) coefficients of Vergados 
\cite{Vergados}. 

The final result in the contraction limit reads 
\begin{eqnarray}
\langle [N], (2N,0), \tilde\chi=0, L || ((d^\dagger s)^{(2)}Q^{(2)})^{(2)} ||  [N], (2N,0), \tilde\chi=0, L \rangle = \nonumber \\
-{1\over 14 \sqrt{2}}  N \sqrt{2L+1} 
\left[ {2\over 3} {(2N-L)(2N+L+1)\over (2N-1)(2N)} {(2L-3)^2 (2L+5)^2 \over (2L-1)^2 (2L+3)^2} \right. \nonumber\\
+{(2N-L-2)\sqrt{(2N-L)(2N+L+3)} \over (2N-1)(2N)} {\sqrt{(2L-1)(2L)(2L+2)(2L+4)(2L+6)(2L+7)} \over (2L+1)^2 (2L+3)^2} \nonumber \\
\left. +{(2N+L-1)\sqrt{(2N-L-2)(2N+L+1)} \over (2N-1)(2N)} {\sqrt{(2L-5)(2L-4)(2L-2)(2L)(2L+2)(2L+3)} \over (2L+1) (2L-1)^2} \right].
\end{eqnarray} 

In the same way in the contraction limit one finds 
\begin{eqnarray}
\langle [N], (2N,0), \tilde\chi=0, L || (Q^{(2)} (d^\dagger s)^{(2)})^{(2)}  ||  [N], (2N,0), \tilde\chi=0, L \rangle = \nonumber \\
-{1\over 14 \sqrt{2}}  N \sqrt{2L+1}
\left[ {2\over 3} {(2N-L)(2N+L+1)\over (2N-1)(2N)} {(2L-3)^2 (2L+5)^2 \over (2L-1)^2 (2L+3)^2} \right. \nonumber\\
+{(2N+L+1)\sqrt{(2N-L)(2N+L+3)} \over (2N-1)(2N)} {(2L-1)(2L)(2L+4)\over (2L+1) (2L+3)^2} \nonumber \\
\left. +{(2N-L)\sqrt{(2N-L+2)(2N+L+1)} \over (2N-1)(2N)} {(2L-2)(2L+2)(2L+3)\over (2L+1) (2L-1)^2} \right]. \nonumber\\
\end{eqnarray}

For the terms involving $s^\dagger \tilde d$ in the same way one finds 
\begin{eqnarray}
\langle [N], (2N,0), \tilde\chi=0, L || ((s^\dagger \tilde d)^{(2)}Q^{(2)})^{(2)} ||  [N], (2N,0), \tilde\chi=0, L \rangle = \nonumber \\
-\langle [N], (2N,0), \tilde\chi=0, L || (Q^{(2)} (d^\dagger s)^{(2)})^{(2)}  ||  [N], (2N,0), \tilde\chi=0, L \rangle, \\
\langle [N], (2N,0), \tilde\chi=0, L || (Q^{(2)} (s^\dagger \tilde d)^{(2)})^{(2)}  ||  [N], (2N,0), \tilde\chi=0, L \rangle = \nonumber \\
-\langle [N], (2N,0), \tilde\chi=0, L || ((d^\dagger s)^{(2)}Q^{(2)})^{(2)} ||  [N], (2N,0), \tilde\chi=0, L \rangle .
\end{eqnarray}

Using these results in the calculation of the matrix elements of $[H_2,\hat Q^{(2)}_{SU(3)}]$  [including only the terms appearing 
in the second line of Eq. (\ref{interm4}), with $\hat Q^{(2)}_\chi$ replaced by $\hat Q^{(2)}_{SU(3)}$], we find Eq. (\ref{meH_2}). 

\section*{Appendix 4. The O(6)$\to$ [R$^5]$SO(5) contraction}

A procedure similar to that of Appendix 2 is followed in the contraction of O(6) to [R$^5]$SO(5) \cite{MtV84,Elliott171}.
This is a procedure in which the full O(6) algebra, consisting of 15 noncommuting generators, is shrinked
into an SO(5) algebra (consisting of 10 noncommuting generators), accompanied by 5 mutually commuting 
operators (the quadrupole operators). The resulting  algebraic structure is known \cite{Elliott171} to be the algebra 
of the $\gamma$-unstable rotator. 

The commutation relation for the quadrupole operators reads 
\begin{equation}
[\hat Q^{(2)}_{O(6),\xi}, \hat Q^{(2)}_{O(6),\nu} ]= 2 \sum_{k=1,3} (2 \xi 2 \nu | k \xi+\nu) (d^\dagger \tilde d)^{(k)}_{\xi+\nu}.
\end{equation}
The second order Casimir operator is \cite{IA}  
\begin{equation}\label{O6C}
\hat C_2 [O(6)]= 2 \hat Q^{(2)}_{O(6)} \cdot \hat Q^{(2)}_{O(6)} 
+ 4 \sum_{k=1,3} (d^\dagger \tilde d)^{(k)} \cdot (d^\dagger \tilde d)^{(k)}. 
\end{equation}
Its eigenvalues are 
\begin{equation}\label{O6c}
C_2(\sigma)= 2 \sigma (\sigma+4), 
\end{equation}
where $\sigma$ is the quantum number characterizing the irreps of O(6).  

If we consider O(6) irreps with large $\sigma$,
we can rescale the quadrupole operator as 
\begin{equation}
\hat q^{(2)}_{O(6),\xi}= { \hat Q^{(2)}_{O(6)),\xi} \over \sqrt{C_2(\sigma) } }.
\end{equation}
Then the commutation relation for the quadrupole operators becomes 
\begin{equation}\label{qqO6}
[\hat q^{(2)}_{O(6),\xi}, \hat q^{(2)}_{O(6),\nu}] = 2 \sum_{k=1,3} (2 \xi 2 \nu | k \xi+\nu) { (d^\dagger \tilde d)^{(k)}_{\xi+\nu}
\over C_2(\sigma)}.
\end{equation}
Then in the limit of large $\sigma$ (and small $\tau$, where $\tau$ is the quantum number characterizing 
the irreps of O(5)~) \cite{Elliott171} one gets 
\begin{equation}\label{qqO6b}
[\hat q^{(2)}_{O(6),\xi}, \hat q^{(2)}_{O(6),\nu}] = 0.
\end{equation}
This procedure is called the contraction of O(6) to [R$^5]$SO(5), where [R$^5]$SO(5) is the algebra 
of the $\gamma$-unstable rotator, generated by the operators of SO(5) and the five commuting 
operators $q^{(2)}_{O(6),\xi}$, $\xi=-2,-1,0,1,2$, which are the coordinates \cite{Elliott171}.

An immediate consequence of Eqs. (\ref{qqO6}) and (\ref{qqO6b}) is that, in the contraction limit, terms proportional to 
$(d^\dagger \tilde d)^{k}$  can be ignored. 

The most leading O(6) irrep, to which the ground state band belongs, is $(N)$.
Thus in the large boson number limit it is appropriate to use this contraction. 
The intrinsic quadrupole moment will then be 
\begin{equation}
q'_0= \sigma,
\end{equation}
as can be seen from Eqs. (\ref{O6C}) and (\ref{O6c}). 
Thus in the case of interest the intrinsic quadrupole moment becomes 
\begin{equation}
q'_0= N. 
\end{equation}
It should be noticed that the above results have been obtained in irreps with $\sigma >> \tau$, 
thus they regard the low lying part of the spectrum (since $L\leq 2\tau$, as seen from the algorithm 
of the SO(5)$\supset$SO(3) reduction \cite{IA}).




\begin{table}

\caption{Single boson matrix elements derived according to Ref. \cite{Rosensteel}.
See Appendix 3 for further discussion. 
}

\bigskip

\begin{tabular}{ l  }
\hline
\\
$\langle [N], (2N,0), \tilde\chi=0, L || s^\dagger ||  [N-1], (2N-2,0), \tilde\chi=0, L \rangle
=\sqrt{(2N-L)(2N+L+1)\over 3(2N-1)(2N)} \sqrt{N} \sqrt{2L+1}$ \\
\\
$\langle [N], (2N,0), \tilde\chi=0, L || d^\dagger ||  [N-1], (2N-2,0), \tilde\chi=0, L \rangle
=-\sqrt{2(2N-L)(2N+L+1)L(L+1)\over 3(2N-1)(2N)(2L-1)(2L+3)} \sqrt{N} \sqrt{2L+1} $\\
\\
$\langle [N], (2N,0), \tilde\chi=0, L || d^\dagger ||  [N-1], (2N-2,0), \tilde\chi=0, L+2 \rangle
=\sqrt{(2N-L-2)(2N-L)(L+2)(L+1)\over (2N-1)(2N)(2L+1)(2L+3)} \sqrt{N} \sqrt{2L+1} $\\
\\
$\langle [N], (2N,0), \tilde\chi=0, L || d^\dagger ||  [N-1], (2N-2,0), \tilde\chi=0, L-2 \rangle
=\sqrt{(2N+L-1)(2N+L+1)(L-1)L\over (2N-1)(2N)(2L-1)(2L+1)} \sqrt{N} \sqrt{2L+1} $\\
\\
$\langle [N-1], (2N-2,0), \tilde\chi=0, L || s ||  [N], (2N,0), \tilde\chi=0, L \rangle
= -\sqrt{(2N-L)(2N+L+1)\over 3(2N-1)(2N)} \sqrt{N} \sqrt{2L+1} $\\
\\
$\langle [N-1], (2N-2,0), \tilde\chi=0, L || \tilde d ||  [N], (2N,0), \tilde\chi=0, L \rangle
= -\sqrt{2(2N-L)(2N+L+1)L(L+1)\over 3(2N-1)(2N)(2L-1)(2L+3)} \sqrt{N} \sqrt{2L+1} $ \\
\\
$\langle [N-1], (2N-2,0), \tilde\chi=0, L || \tilde d ||  [N], (2N,0), \tilde\chi=0, L+2 \rangle
= \sqrt{(2N+L+1)(2N+L+3)(L+1)(L+2)\over (2N-1)(2N)(2L+3)(2L+5)} \sqrt{N} \sqrt{2L+5}$ \\
\\
$\langle [N-1], (2N-2,0), \tilde\chi=0, L || \tilde d ||  [N], (2N,0), \tilde\chi=0, L-2 \rangle
= \sqrt{(2N-L)(2N-L+2)L(L-1)\over (2N-1)(2N)(2L-3)(2L-1)} \sqrt{N} \sqrt{2L-3}$ \\
\\
\hline 
\end{tabular}
\end{table}

\newpage

\begin{figure}
\center{{\includegraphics[height=170mm]{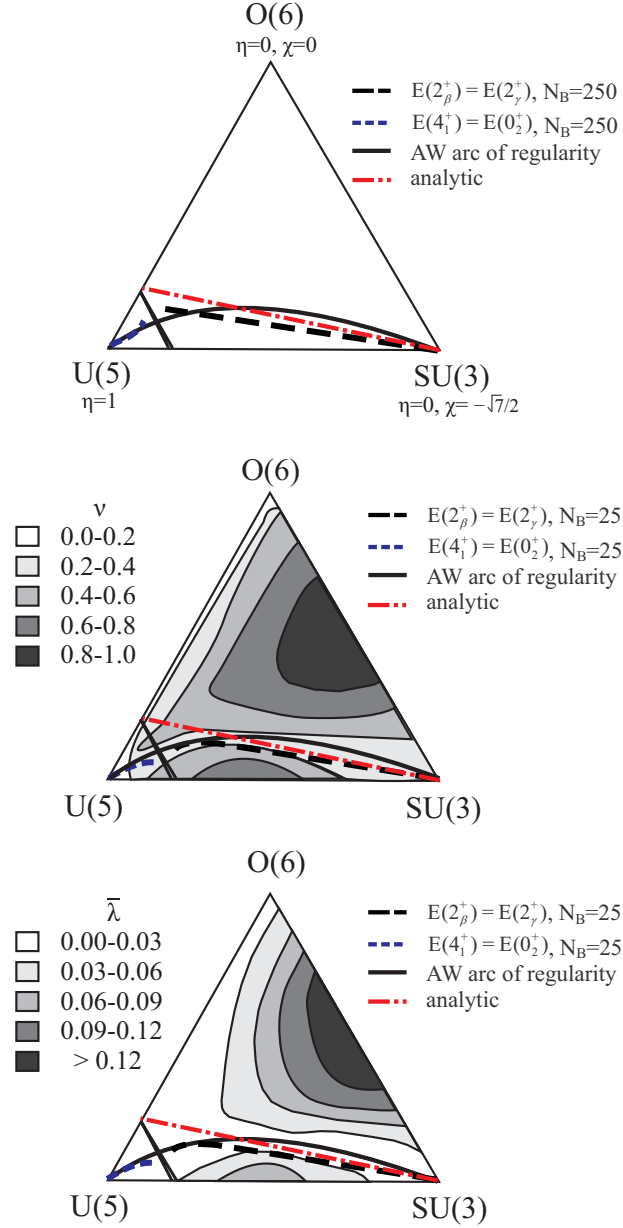}}} 
\caption{(Color online) IBA symmetry triangle in the parametrization of Eq. (\ref{HAW})
with the three dynamical symmetries, the Alhassid--Whelan arc of regularity [Eq. (\ref{arc})], 
and the present line of Eq. (\ref{line_eta}) (labelled as analytic).  
The shape coexistence region \cite{IZC} between spherical and deformed phases is shown by slanted lines
near the U(5) vertex. In addition, the loci of the degeneracies $E(2_\beta^+)$=$E(2_\gamma^+)$ 
(dashed line on the right, corresponding to the SU(3) QDS discussed in Ref. \cite{arcPRL}) 
and  $E(4_1^+)$=$E(0_2^+)$ (dotted line on the left, also discussed in Ref. \cite{arcPRL}) are shown for $N_B$=250 (top) and
$N_B$ = 25 (bottom).  In the middle and bottom parts, the $\nu$-diagram and the $\bar \lambda$-diagram, 
based on Ref. \cite{AWNPA}, are shown. See subsection 3.1 for further discussion. }
\end{figure}

\begin{figure}
\center{{\includegraphics[height=60mm]{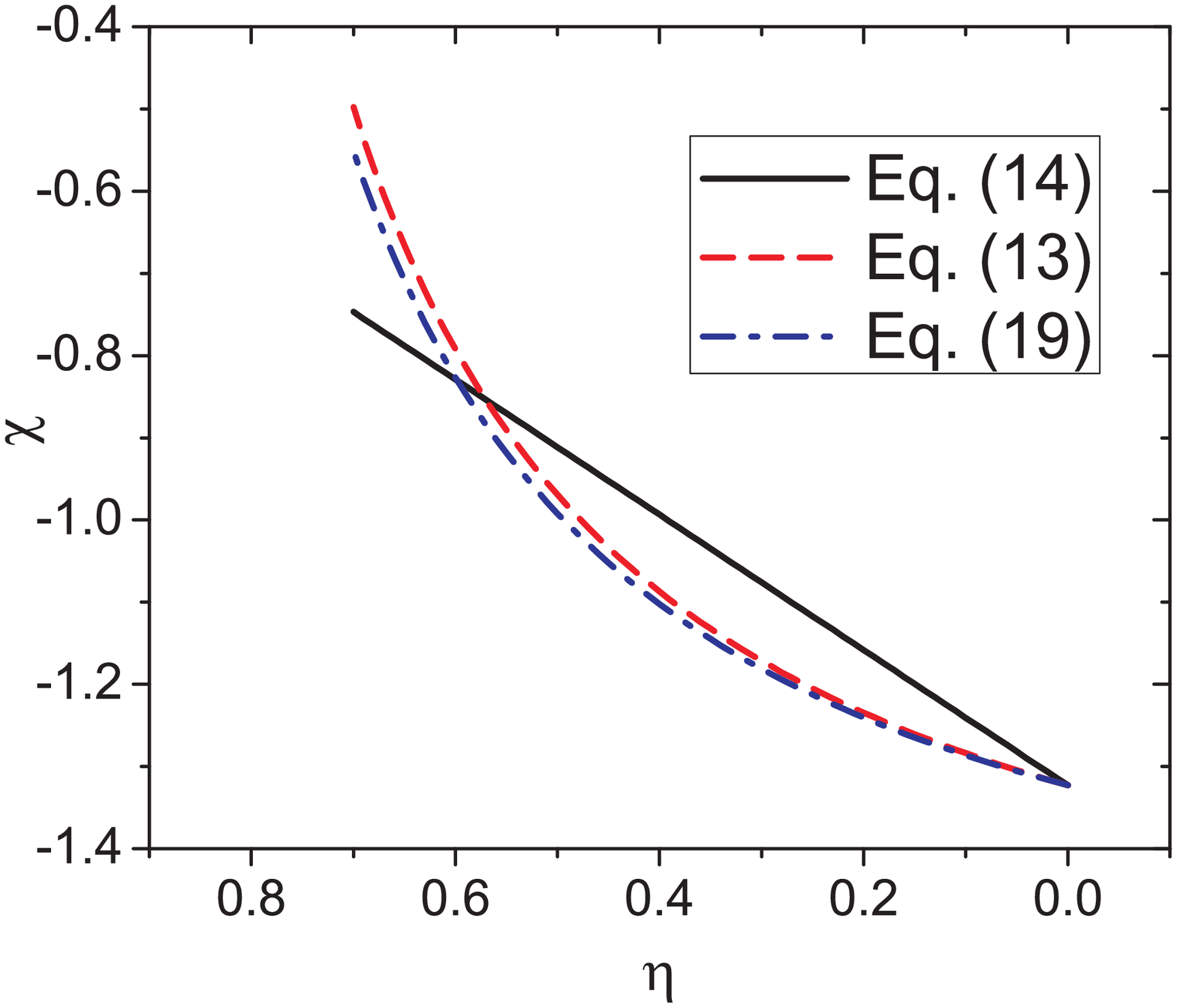}}}
\caption{(Color online) Location of the arc of regularity, as described by the original Eq. (\ref{arc}),
and as predicted by the findings of the present work, Eqs. (\ref{line_eta}) and (\ref{gsb_AW}). 
The $\eta$ axis has been reversed, in order to correspond directly to Fig.~1.   
See subsections 3.1 and 3.2 for further discussion. }
\end{figure}


\begin{thebibliography}{99}

\bibitem{IA}
F. Iachello and A. Arima, {\it The Interacting Boson Model} (Cambridge 
University Press, Cambridge, 1987). 

\bibitem{AWPRL}
Y. Alhassid and N. Whelan, Phys. Rev. Lett. {\bf 67}, 816 (1991).

\bibitem{AWNPA}
N. Whelan and Y. Alhassid, Nucl. Phys. A {\bf 556}, 42 (1993). 

\bibitem{triangle}
R. F. Casten, {\it Nuclear Structure from a Simple Perspective}
(Oxford University Press, Oxford, 1990).

\bibitem{Talmi}
A. Leviatan, A. Novoselsky, and I. Talmi, Phys. Lett. B {\bf 172}, 144 (1986). 

\bibitem{CJPRE}
P. Cejnar and J. Jolie, Phys. Rev. E {\bf 58}, 387 (1998). 

\bibitem{AWexp}
J. Jolie, R. F. Casten, P. Cejnar, S. Heinze, E. A. McCutchan, and N. V. Zamfir, 
Phys. Rev. Lett. {\bf 93}, 132501 (2004). 

\bibitem{Macek75}
M. Macek, P. Str\'ansk\'y, P. Cejnar, S. Heinze, J. Jolie, and J. Dobe\v s, 
Phys. Rev. C {\bf 75}, 064318 (2007).  

\bibitem{arcPRL}
D. Bonatsos, E.A. McCutchan, and R.F. Casten, Phys. Rev. Lett. {\bf 104}, 022502 (2010). 

\bibitem{IZC}
F. Iachello, N. V. Zamfir, and R. F. Casten, Phys. Rev. Lett. {\bf 81}, 1191 (1998).

\bibitem{Rowe1225}
D. J. Rowe, Phys. Rev. Lett. {\bf 93}, 122502 (2004). 

\bibitem{Rowe2325}
D. J. Rowe,P. S. Turner, and G. Rosensteel, Phys. Rev. Lett. {\bf 93}, 232502 (2004). 

\bibitem{Rowe745}
D. J. Rowe, Nucl. Phys. A {\bf 745}, 47 (2004). 

\bibitem{Rowe756}
P. S. Turner and D. J. Rowe, Nucl. Phys. A {\bf 756}, 333 (2005). 

\bibitem{Rowe759}
G. Rosensteel and D. J. Rowe, Nucl. Phys. A {\bf 759}, 92 (2005). 

\bibitem{Macek80}
M. Macek, J. Dobe\v{s}, and P. Cejnar, Phys. Rev. C {\bf 80}, 014319 (2009).

\bibitem{Macek82}
M. Macek, J. Dobe\v{s}, and P. Cejnar, Phys. Rev. C {\bf 82}, 014308 (2010).

\bibitem{Inonu}
E. \.{I}n\H{o}n\H{u} and E. P. Wigner, Proc. Natl. Acad. Sci. (N.Y.) {\bf 39}, 510 (1953). 

\bibitem{LeBlanc}
R. Le Blanc, J. Carvalho, and D. J. Rowe, Phys. Lett. B {\bf 140}, 155 (1984). 

\bibitem{RowePPNP}
D. J. Rowe, Prog. Part. Nucl. Phys. {\bf 37}, 265 (1996). 

\bibitem{Ui}
H. Ui, Prog. Theor. Phys. {\bf 44}, 153 (1970). 

\bibitem{MtV84}
J. Meyer-ter-Vehn, Phys. Lett. B {\bf 84}, 10 (1979). 

\bibitem{Elliott171}
J. P. Elliott, P. Park, and J. A. Evans, Phys. Lett. B {\bf 171}, 145 (1986). 

\bibitem{Werner}
V. Werner, P. von Brentano, R. F. Casten, and J. Jolie, Phys. Lett. B {\bf 527}, 55 (2002).

\bibitem{Deans}
D. H. Feng, R. Gilmore, and S. R. Deans, Phys. Rev. {\bf C 23}, 1254 (1981). 

\bibitem{Libby}
E. A. McCutchan, D. Bonatsos, and N. V. Zamfir, Phys. Rev. C {\bf 74}, 034306 (2006). 

\bibitem{Vergados}
J. D. Vergados, Nucl. Phys. A {\bf 111}, 681 (1968). 

\bibitem{Zamfir66}
N. V. Zamfir, P. von Brentano, R. F. Casten, and J. Jolie, Phys. Rev. C {\bf 66}, 021304 (2002).

\bibitem{AL25}
Y. Alhassid and A. Leviatan, J. Phys. A: Math. Gen. {\bf 25}, L1265 (1992).

\bibitem{Lev77}
A. Leviatan, Phys. Rev. Lett. {\bf 77}, 818 (1996).

\bibitem{LVI89}
A. Leviatan and P. Van Isacker, Phys. Rev. Lett. {\bf 89}, 222501 (2002).

\bibitem{Lev2010}
A. Leviatan, arXiv:1004.5325v1 [nucl-th] (2010). 

\bibitem{Kota}
Y.  D. Devi and V. K. B. Kota, Pramana-J. Phys. {\bf 39}, 413 (1992). 

\bibitem{EINPA}
J. Engel and F. Iachello, Nucl. Phys. A {\bf 472}, 61 (1987). 

\bibitem{Kusn1}
D. Kusnezov, J. Phys. A: Math. Gen. {\bf 22}, 4271 (1989).  

\bibitem{Kusn2}
D. Kusnezov, J. Phys. A: Math. Gen. {\bf 23}, 5673 (1990). 

\bibitem{Lev179}
A. Leviatan, Ann. Phys. (N.Y.) {\bf 179}, 201 (1987). 

\bibitem{WAL}
N. Whelan, Y. Alhassid, and A. Leviatan, Phys. Rev. Lett. {\bf 71}, 2208 (1993). 

\bibitem{AriasPRL}
J. M. Arias, J. E. Garc\'{\i}a-Ramos, and J. Dukelsky, Phys. Rev. Lett. {\bf 93}, 212501 (2004).

\bibitem{CIPRL}
M. A. Caprio and F. Iachello, Phys. Rev. Lett. {\bf 93}, 242502 (2004).

\bibitem{CIAP}
M. A. Caprio and F. Iachello, Ann. Phys. (N.Y.) {\bf 318}, 454 (2005). 

\bibitem{PVI}
P. Van Isacker, A. Bouldjedri, and S. Zerguine, Nucl. Phys. A {\bf 836}, 225 (2010). 

\bibitem{IBAR1}
R. J. Casperson, IBAR code (unpublished). 

\bibitem{IBAR2}
E. Williams, R. J. Casperson, and V. Werner, Phys. Rev. C {\bf 77}, 061302(R) (2008).

\bibitem{Edmonds}
A. R. Edmonds, {\it Angular Momentum in Quantum Mechanics} (Princeton University Press,
Princeton, 1957). 

\bibitem{Vassanji}
D. J. Rowe, M. G. Vassanji, and J. Carvalho, Nucl. Phys. A {\bf 504}, 76 (1989). 

\bibitem{ElliottII}
J. P. Elliott, Proc. Roy. Soc. Ser. A {\bf 245}, 562 (1958). 

\bibitem{Rosensteel}
G. Rosensteel, Phys. Rev. C {\bf 41}, 730 (1990). 


\end{thebibliography}
\end{document}